\documentclass
[aps,prl,amsfonts,amssymb,twocolumn,amsmath,preprintnumbers,nofootinbib,floatfix,superscriptaddress,longbibliography]{revtex4-2}%
\usepackage[dvips]{graphics}
\usepackage{graphicx}
\usepackage{bm}
\usepackage{amsmath}
\usepackage{amsfonts}
\usepackage{amssymb}
\usepackage{xcolor}
\usepackage{subfigure}
\usepackage{tabularx}
\usepackage{multirow}
\usepackage{braket}
\usepackage{commath}
\usepackage[colorlinks,linkcolor=blue,anchorcolor=blue,citecolor=blue,urlcolor=blue]%
{hyperref}%
\providecommand{\U}[1]{\protect\rule{.1in}{.1in}}

\begin{document}

\title{Orbital Chern Insulator at $\nu=-2$ in Twisted MoTe$_{2}$}
\author{Feng-Ren Fan}
\thanks{These authors contributed equally to this work.}
\affiliation{Department of Physics, The University of Hong Kong, Hong Kong, China}
\affiliation{HKU-UCAS Joint Institute of Theoretical and Computational Physics at Hong Kong, China}
\author{Cong Xiao}
\thanks{These authors contributed equally to this work.}
\affiliation{Institute of Applied Physics and Materials Engineering, University of Macau, Taipa, Macau, China}
\affiliation{Department of Physics, The University of Hong Kong, Hong Kong, China}
\affiliation{HKU-UCAS Joint Institute of Theoretical and Computational Physics at Hong Kong, China}
\author{Wang Yao}
\email[]{wangyao@hku.hk}
\affiliation{Department of Physics, The University of Hong Kong, Hong Kong, China}
\affiliation{HKU-UCAS Joint Institute of Theoretical and Computational Physics at Hong Kong, China}

\begin{abstract}
In twisted MoTe$_{2}$, latest transport measurement has reported observation
of quantum anomalous Hall effect at hole filling $\nu=-1$, which undergoes a
topological phase transition to a trivial ferromagnet as layer hybridization
gets suppressed by interlayer bias $D$.
Here we show that this underlies the existence of an orbital Chern insulating state with
gate ($D$) switchable sign in an antiferromagtic spin background at hole filling
$\nu=-2$. From momentum-space Hartree Fock calculations, we find this state
has a topological phase diagram complementary to that of the $\nu=-1$ one: by
sweeping $D$ from negative to positive, the Chern number of this $\nu=-2$
state can be switched between $+1$, $0$, and $-1$, accompanied by a sign
change of a sizable orbital magnetization. In range of $D$ where this
antiferronagnet is the ground state, the orbital magnetization allows magnetic
field initialization of the spin antiferromagnetic order and the Chern number.
\end{abstract}

\maketitle

\begin{figure*}[ptbh]
\includegraphics[width=0.9\textwidth]{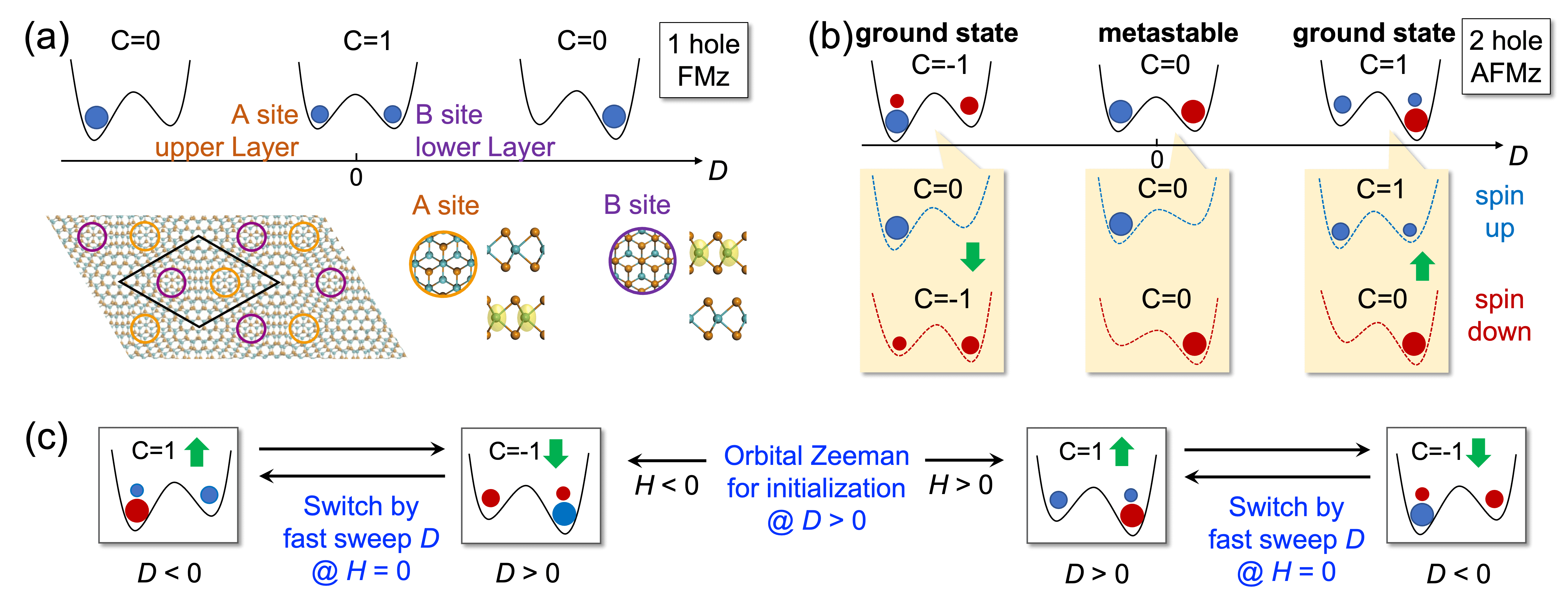}
\caption{(a) Lower panel: layer-sublattice locked honeycomb superlattice in R stacking tMoTe$_2$. The side and top views of the local stacking registries are shown for the A and B moir\'{e} sites, where holes are localized in lower and upper layer respectively. Upper panel: black curves illustrates the potential landscape for the two moir\'{e} sites. At filling of 1 hole per moir\'{e} cell ($\nu=-1$), the ground state is an out-of-plane ferromagnet (FMz).
When the interlayer bias $D$ is small, the FMz state is layer hybridized and exhibits the quantum anomalous Hall (QAH) effect. When $|D|$ exceeds a critical value, the FMz state becomes topologically trivial as the layer hybridization is suppressed (c.f. Fig.~\ref{fig:filling1}). 
C denotes Chern number.
(b) A complementary topological phase transition of
an out-of-plane antiferromagnetic insulating state (AFMz) at $\nu=-2$ (c.f.
Fig.~\ref{fig:altermagnet} \& \ref{fig:orbinsulator}). Insets illustrate the
layer wavefunctions, effective potentials (moir\'{e} + Hartree, dashed curves),
and Chern numbers for spin up and down carriers separately. (c) In
range of $D$ where the AFMz QAH is ground state, its orbital magnetization
(green arrows) allows initialization of the AFMz configurations by magnetic
field $H$. The AFMz QAH can then be electrically switched at zero $H$, by
sweeping $D$ at time scale faster than spin relaxation.}%
\label{fig:schematic}%
\end{figure*}

Bilayer MoTe$_{2}$ is a direct-gap semiconductor~\cite{lezama_indirect--direct_2015},
 while under twisted rhombohedral (R) stacking it hosts a honeycomb moir\'{e} superlattice with the
two triangular sublattices residing on opposite layers~\cite{wu_topological_2019, yu_giant_2020}.
From such a layer
pseudospin texture, carrier motion acquires Berry phases that realize fluxed
superlattice~\cite{yu_giant_2020, zhai_theory_2020},
 which underlies the Kane-Mele type topological dispersion in the lowest energy
minibands~\cite{wu_topological_2019, devakul_magic_2021}.
With intrinsic ferromagnetism arising from the direct Coulomb
exchange between moir\'{e} orbitals~\cite{anderson_programming_2023},
 such a layer-sublattice locked moir\'{e}
superlattice has emerged as an exciting platform for investigating the quantum
anomalous Hall (QAH) effect~\cite{cai_signatures_2023, zeng_thermodynamic_2023, park_observation_2023, foutty_mapping_2023, li_spontaneous_2021, xu_observation_2023}.
Thermodynamic evidences of both  integer and fractional QAH effects
 were observed at hole filling of $\nu = -1$, and $\nu=-2/3, -3/5$  respectively
 in twisted bilayer MoTe$_2$ (tMoTe$_2$)~\cite{cai_signatures_2023, zeng_thermodynamic_2023}.
Most excitingly, with a novel contact gate design on tMoTe$_2$,
 direct observations of both integer and fractional QAH effects are reported in latest transport measurements~\cite{park_observation_2023}.
The experiments further demonstrate an electrically controlled topological phase transition at
$\nu=-1$ from QAH to a trivial ferromagnet as layer hybridization gets
suppressed by interlayer bias (Fig.~\ref{fig:schematic}(a))~\cite{park_observation_2023}.
These QAH transports are subsequently reproduced in an independent experiment~\cite{xu_observation_2023}.

Despite the prevailing notion that a net spin magnetization is essential for
the emergence of QAH, it has been shown that the phenomenon can also be
realized in collinear antiferromagnets without net spin magnetization~\cite{jiang_antiferromagnetic_2018, dai_quantum_2022, guo_quantum_2023}.
In particular, a Hartree-Fock mean field calculation of the Kane-Mele Hubbard
model has found a QAH phase at filling factor 2~\cite{jiang_antiferromagnetic_2018}
 under an inversion-breaking
ionic potential, which is an out-of-plane N\'{e}el antiferromagnet with zero total
magnetization. This type of QAH state is reminiscent of the anomalous Hall
effect in metallic antiferromagnets where the absence of combined time
reversal and space inversion symmetry leads to spin splitting in Bloch bands~\cite{hayami_momentum-dependent_2019,shao2021spin,smejkal_beyond_2022, smejkal_emerging_2022, mazin_editorial_2022, smejkal_anomalous_2022, feng_anomalous_2022, gonzalez_betancourt_spontaneous_2023}.
QAH effect in such phase may potentially integrate
topological electronics with antiferromagnetic spintronics. However, there
always exist two degenerate QAH states with opposite N\'{e}el order and Chern
number, which are difficult to distinguish from the fully compensated spin
magnetization, and their domain structures also hinder the observation and
utilization of QAH. Alternatively, QAH can also be associated with
magnetization of purely orbital nature, as indicated by the St{\v{r}}eda formula~\cite{streda_theory_1982}
between orbital magnetization and Chern number~\cite{zhu_voltage-controlled_2020}.
Such orbital Chern insulator has been observed recently in magic angle twisted bilayer graphene~\cite{sharpe_emergent_2019, serlin_intrinsic_2020, polshyn_electrical_2020},
whereas at modest magnetic field evidence of the fractional Chern insulator
has also been reported~\cite{xie_fractional_2021}.

\begin{figure}[ptbh]
\includegraphics[width=0.48\textwidth]{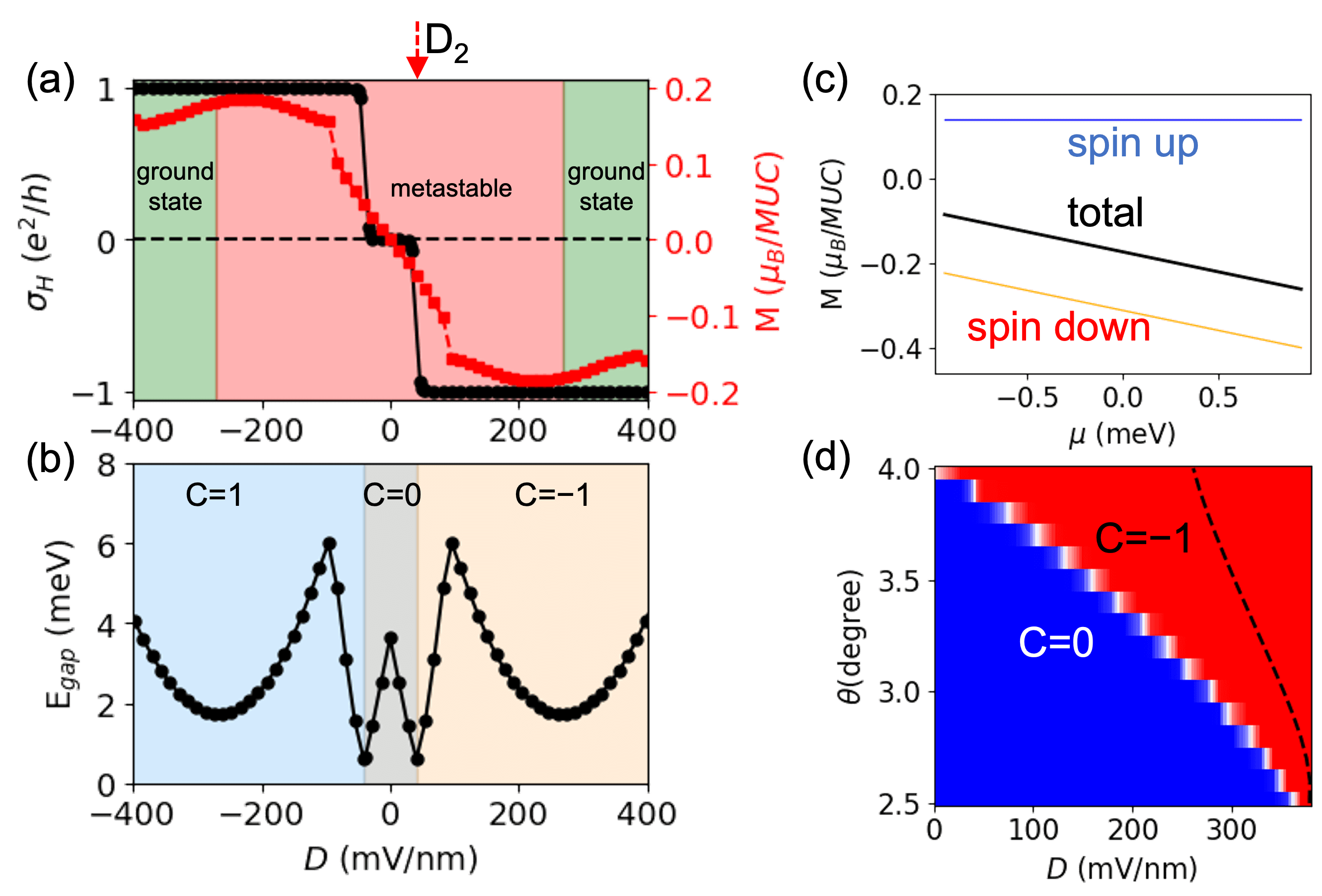}
\caption{ (a) The anomalous Hall conductivity (black circles) and orbital
magnetization (red squares) as functions of interlayer bias in the $\nu=-2$ AFMz state with twist
angle $\theta=3.9^{\circ}$. 
The background color denotes region where AFMz is the ground state (green) or metastable (red).
(b) The band gap of AFMz state as a function of interlayer bias. Gap closing
occurs at the transition points.
(c) Orbital magnetization of the $\nu = -2$ AFMz state,
 under interlayer bias $D=300~$mV/nm, as a function of chemical potential $\mu$ that varies inside the gap.
The blue and orange lines are the orbital magnetization
contributed by spin-up and spin-down holes, respectively, and the black line
is their sum.
(d) Topological phase diagram
of the $\nu = -2$ AFMz state, as a function of twist angle and interlayer bias.
The AFMz state is the ground state to the right side of black dashed curve, and is metastable to the left.}%
\label{fig:orbinsulator}%
\end{figure}

Here we show that the experimentally observed QAH state at $\nu=-1$ filling
tMoTe$_{2}$ underlies the existence of an orbital Chern
insulator under finite interlayer bias $D$ at $\nu=-2$, where the layer
hybridization is key to the miniband topology in the layer-sublattice locked
moir\'{e}. A unified momentum-space Hartree Fock calculation reproduces the
observed continuous topological phase transition of the out-of-plane
ferromagnet at $\nu=-1$ (FMz)~\cite{park_observation_2023},
 and finds this $\nu=-2$ state featuring
out-of-plane N\'{e}el-type spin order (AFMz) with spin split minibands and sizable
orbital magnetization. Its topological phase diagram as function of $D$ is
essentially complementary to that of the $\nu=-1$ state
(Fig.~\ref{fig:schematic}(b)). At small $|D|$, either spin component occupies
a layer-polarized and topologically trivial miniband, while for $\left\vert
D\right\vert >D_{2}$, the system becomes a Chern insulator (c.f. Fig.~\ref{fig:orbinsulator}), as a
pronounced layer-hybridization is restored for one spin component due to the
balance between the interlayer bias $D$ and the Hartree potential from the
other spin. By sweeping $D$ from positive to negative, the Chern number is
switched between $+1$, $0$, and $-1$, accompanied by a sign change of orbital
magnetization. In range of $D$ where this AFMz state is ground state, the
sizable orbital magnetization (O(0.01) $\mu_{\text{B}}/$nm$^{2}$) makes
possible initialization of the spin N\'{e}el order and the Chern number in applied
magnetic field. And the Chern number can then be electrically switched sign at
zero magnetic field, by sweeping $D$ at time scale faster than magnetic
relaxation (Fig.~\ref{fig:schematic}(c)). These findings imply new
possibilities for antiferromagnetic spintronics and topological electronics based on
layer-sublattice locked moir\'{e} in twisted transition metal dichalcogenides (TMDs).

\emph{{\color{blue} Method.}}--To describe the spontaneous time reversal
symmetry breaking from electron-electron interactions, we employ
self-consistent momentum (\textit{k})-space Hartree Fock calculations based on
the continuum model for twisted R stacking homobilayer TMDs. The Hartree-Fock
mean field Hamiltonian reads%
\begin{equation}
H_{HF}=H_{0}+H_{int},\label{eq:hmthf}%
\end{equation}
where $H_{0}$ is the single-particle continuum Hamiltonian~\cite{wu_topological_2019}.
\begin{equation}
H_{0}=\left(
\begin{array}
[c]{cc}%
h_{k}^{t}-\frac{E_{D}}{2} & w\\
w^{\dagger} & h_{k}^{b}+\frac{E_{D}}{2}%
\end{array}
\right)  ,\label{eq:hmt0}%
\end{equation}
where $h_{k}^{\alpha}=-\frac{\left(  \boldsymbol{k}-\boldsymbol{K}^{\alpha
}\right)  ^{2}}{2m_{\alpha}}+V_{k}^{\alpha}$ is the single-layer Hamiltonian
near valley $\boldsymbol{K}^{\alpha}$ in layer $\alpha$, $E_{D}=Dd_{0}%
/\epsilon$ is the energy difference caused by the interlayer bias $D$, with
$d_{0}$ and $\epsilon$ being respectively the interlayer distance and relative
dielectric constant, $V_{k}^{\alpha}$ is the Fourier transformation of the
intralayer moir\'{e} potential $V^{\alpha}(\boldsymbol{r})=-2V_{0}%
\sum_{i=1,3,5}\cos\left(  \boldsymbol{g}_{i}\cdot\boldsymbol{r}+\phi_{\alpha
}\right)  $, and $w(\boldsymbol{r})=w_{0}\left(  1+e^{i\boldsymbol{g}_{2}%
\cdot\boldsymbol{r}}+e^{i\boldsymbol{g}_{3}\cdot\boldsymbol{r}}\right)  $ is
ther interlayer tunneling. $\boldsymbol{g}_{i}$ is the first-shell moir\'{e}
reciprocal vectors. The model parameters for tMoTe$_{2}$ are
$m_{t}=m_{b}=-0.62m_{e}$~\cite{wu_topological_2019},
 $V_{0}=20.82~$meV, $\phi_{t}=-\phi_{b}%
=-107.7^{\circ}$, and $w=-23.8~$meV~\cite{wang_fractional_2023}.
$H_{int}$ in Eq.~(\ref{eq:hmthf}) is the
interaction term which, based on the plane-wave basis expanded by the
moir\'{e} reciprocal lattice vectors, reads
\begin{equation}
H_{\text{int }}=\frac{1}{2\Omega}\sum_{\substack{k,k^{\prime},q,\alpha
\beta,\tau\tau^{\prime}}}V_{\alpha\beta}(q)c_{k+q,\alpha\tau}^{\dagger
}c_{k^{\prime}-q,\beta\tau^{\prime}}^{\dagger}c_{k^{\prime},\beta\tau^{\prime
}}c_{k,\alpha\tau}.\label{eq:hint}%
\end{equation}
Here 
$\tau$ is the valley index, and
$V_{\alpha\beta}$ is the Coulomb interaction taking the form $V_{\alpha\beta
}(q)=\frac{2\pi e^{2}}{\epsilon q}e^{-qd_{0}(1-\delta_{\alpha\beta})}$, with
$\delta_{\alpha\beta}$ being the Kronecker delta.

\begin{figure}[pthb]
\includegraphics[width=0.48\textwidth]{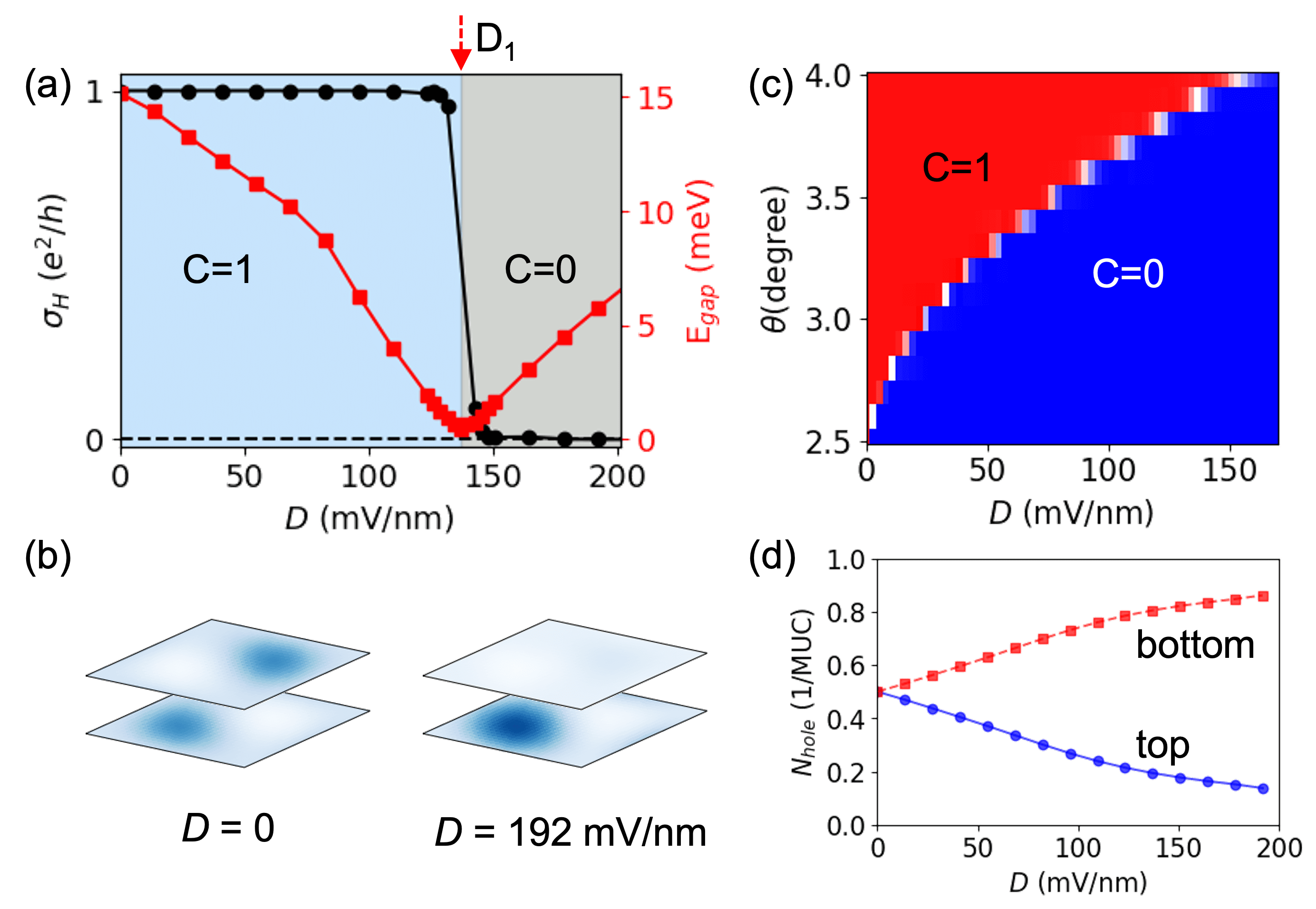}\caption{ (a) The
variation of anomalous Hall conductivity and band gap as functions of
interlayer bias $D$, with twist angle $\theta=3.9^{\circ}$. The black-circle
line represents the Hall conductivity, while the red-square line the band gap.
The background colours denote the Chern number.
$D_1 = 137$~mV/nm is the topological transition point.
(b) The layer distribution of
the doped hole under interlayer bias $D=0$, and $D=192~$mV/nm. (c) Phase
diagram of 1-hole doped tMoTe$_{2}$, as a function of twist angle and
interlayer bias. (d) Number of holes per moir\'{e} unit cell (MUC) on each
layer under different interlayer bias. The blue-circle line denotes the top
layer, whereas the red-square line the bottom layer.}%
\label{fig:filling1}%
\end{figure}

$H_{0}$ features the Kane-Mele type topological dispersion in its lowest two minibands~\cite{wu_topological_2019},
which are predominantly accounted by two moir\'{e} orbitals residing on
opposite layers (c.f. Fig.~\ref{fig:schematic}(a)), forming a honeycomb
superlattice where Berry phase from the layer texture realizes the flux needed
in Kane-Mele model~\cite{yu_giant_2020}.
Earlier work based on the two-orbital Kane-Mele Hubbard
model found that the ground state at filling 2 is a AFMz QAH at a range of
finite ionic potential $\Delta_{AB}$ that lifts the A-B sublattice
(Fig.~\ref{fig:schematic}(a)) degeneracy, which undergoes a magnetic phase
transition with the decrease of $\Delta_{AB}$~\cite{jiang_antiferromagnetic_2018}, becoming a trivial in-plane
antiferromagnet (AFMxy) in the neighbourhood of $\Delta_{AB}=0$~\cite{hohenadler_quantum_2012, qiu_interaction-driven_2023}.
Hartree Fock calculations from Eq.~(\ref{eq:hmthf}) can produce these spin magnetic orders
with qualitatively the same dependence on interlayer bias $D$~\cite{liu_gate-tunable_2023}.

On the other hand, the moir\'{e} orbitals here are rather extended in the
shallow confinement which also varies significantly with $D$, and their direct
Coulomb exchange, absent in the Kane-Mele Hubbard model~\cite{jiang_antiferromagnetic_2018}, is responsible for
the observed FMz QAH at $\nu=-1$~\cite{anderson_programming_2023}.
A \textit{k}-space Hartree-Fock treatment keeping large number of minibands is necessitated to properly account for the
subtle $D$ dependence of the spatial and layer wavefunction, as well as the
orbital magnetization.
The calculations presented below have considered 148 minibands for each spin component.
The orbital magnetization is calculated following ref.~\cite{xiao_berry_2010}.

\begin{figure}[ptbh]
\includegraphics[width=0.48\textwidth]{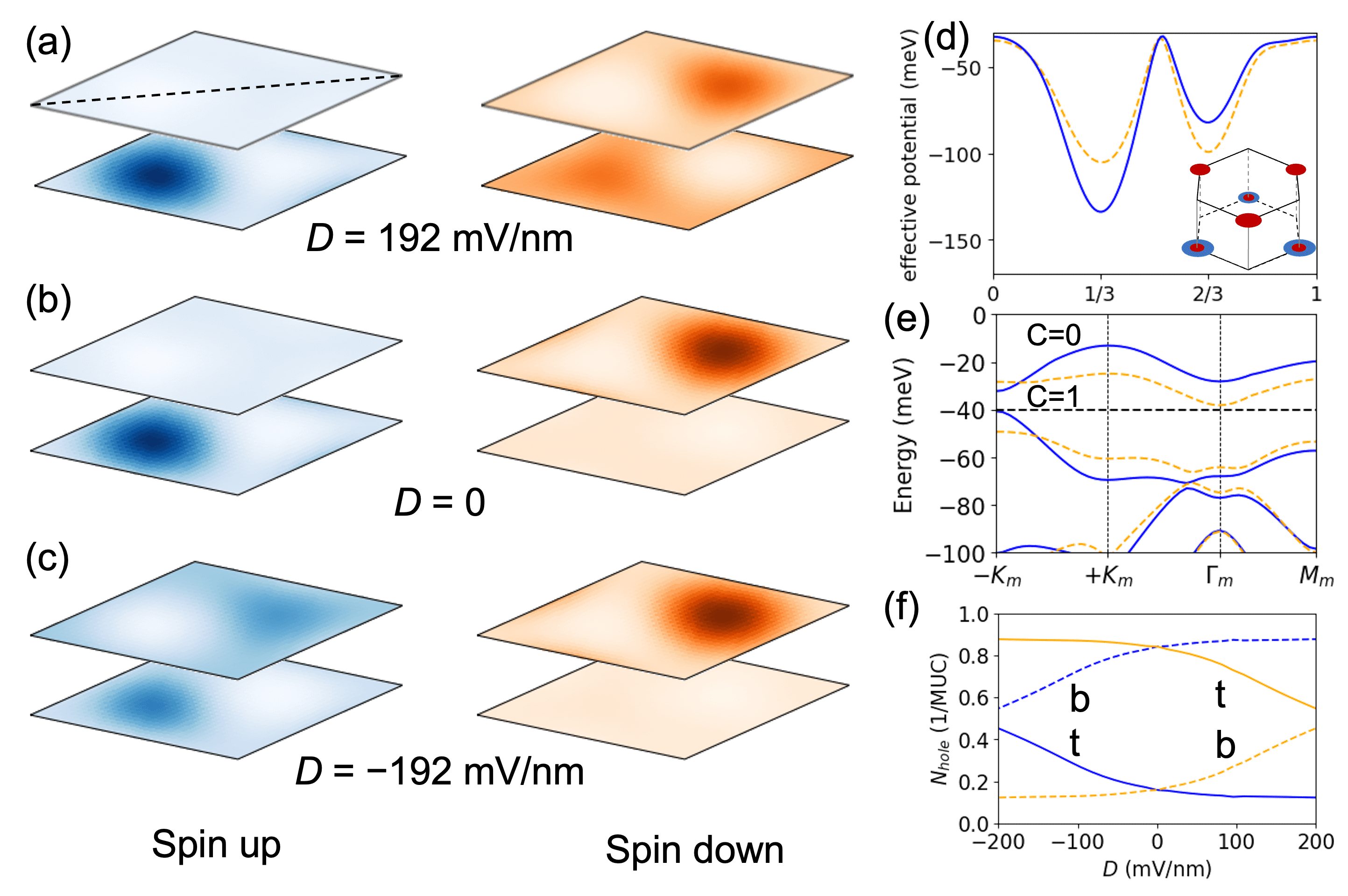}\caption{ (a)-(c)
Layer distribution of wavefunctions for the $\nu=-2$ AFMz states under
interlayer bias $D=192~$mV/nm, 0, and $-192~$mV/nm, respectively.
(d)
Effective potentials (moir\'{e} + Hartree) along the long diagonal of the moir\'{e} cell
 (c.f. part (a)) for spin-up and spin-down holes, respectively, at $D=192~$mV/nm.
Inset is a schematic of magnetic configuration.
Spin-up and spin-down holes are denoted by blue and red circles, respectively, with the size representing the population.
 In (d-f), blue (orange) color denotes spin up (down).
(e) Quasiparticle band dispersion of the $\nu=-2$ AFMz state
 under interlayer bias $D=192~$mV/nm.
(f) Dashed (solid) curves denote number of holes per MUC in top (bottom) layer.}%
\label{fig:altermagnet}%
\end{figure}

\emph{{\color{blue} QAH state at filling $\nu=-1$.}}--We first show the
calculation results at hole filling $\nu=-1$, of which the magnetic and
topological properties have recently been studied in experiment~\cite{anderson_programming_2023, park_observation_2023}.
Figure~\ref{fig:filling1}(a) depicts the variation of anomalous Hall
conductivity of $3.9^{\circ}$ tMoTe$_2$ as a function of interlayer bias. The
Hall conductivity exhibits a jump from $e^{2}/h$ to $0$ at the critical
interlayer bias $D_{1}\sim140~$mV/nm, indicating a topological phase
transition. The band gap, denoted by red squares, varies continuously
 and closes at the transition point. In addition, there is
an inflection point of the band gap near $D\sim75~$mV/nm, which is a result of
the band gap transition from a direct type to an indirect type. All these
features are consistent with the observation of bias driven topological
transition in a recent experiment~\cite{park_observation_2023}.

Such a bias driven topological transition is correlated with the change of
layer wavefunction of the doped hole. Figure~\ref{fig:filling1}(b)
demonstrates the layer distribution of probability density of the doped hole
in the topologically nontrivial state at $D=0$ and the topologically trivial
state at $D=192~$mV/nm. At $D=0$, the distribution of doped hole is
hybridized between the two layers. Because of the layer-sublattice locking, the
doped hole is also equally distributed on sublattices A and B
(Fig.~\ref{fig:schematic}(a)).
In contrast, at $D=192~$mV/nm, the doped hole is
mainly distributed on the bottom layer, being a topologically trivial
system with vanishing Chern number.
Figure~\ref{fig:filling1}(d) shows the variation of the number of holes
on each layer as a function of interlayer bias. As the bias increases, the
system transforms from a fully layer-hybridized state to a strongly
layer-polarized state.

We also explore the twist-angle dependence of the critical bias for the
topological transition by showing the phase diagram in Fig.~\ref{fig:filling1}%
(c). One finds that the critical bias increases with the twist angle. This is
anticipated as the low-energy band width increases with the twist angle,
diminishing the effect of bias tuning.

\emph{{\color{blue}Bias tunable orbital Chern insulator at
$\nu=-2$.}}--The above understanding of filling $\nu=-1$ state enables
exploring the bias controlled topological properties of $\nu=-2$ AFMz state in
tMoTe$_{2}$, as indicated in Fig.~\ref{fig:schematic}.
At $D = 0$, each sublattice hosts a hole with opposite spin, as depicted in
Fig.~\ref{fig:altermagnet}(b).
Each spin species is essentially the same as the
topologically trivial $\nu=-1$ state with strong layer polarization (right
panel of Fig.~\ref{fig:filling1}(b)).
When an interlayer bias is applied,
 one of the spin species starts to be driven to the other layer, becoming more layer-hybridized,
 while the other spin species remains layer-polarized, as shown in
Fig.~\ref{fig:altermagnet}(a) and (c).
The effective potentials (moir\'{e}  + Hartree) for the two spin species at
$D=192~$mV/nm are displayed in Fig.~\ref{fig:altermagnet}(d), where the
potential well for the spin-down hole is balanced by the applied bias and
Hartree potential.
Figure~\ref{fig:altermagnet}(f) illustrates the number of holes on each
layer varying with the bias. When $D$ increases on the positive axis,
 the layer distribution of the spin-up
hole (blue lines) changes little, while the spin-down hole (orange lines)
becomes layer hybridized. Conversely, when $D<0$, the spin-up hole becomes
layer hybridized. In line with the case of $\nu=-1$ state, such
layer-hybridized insulating states are expected to be topologically
nontrivial, which is indeed confirmed by our calculations.
As shown in Fig.~\ref{fig:orbinsulator}(a), when $|D|$ increases, the Chern number jumps
from 0 to 1 or $-1$, depending on the sign of bias. Such jumps signal bias
driven topological phase transitions (Fig.~\ref{fig:orbinsulator}(b)).
Figure~\ref{fig:orbinsulator}(d) shows the topological phase diagram of the
AFMz state as a function of twist angle and interlayer bias. As the twist
angle decreases, the critical bias for the topological phase transition
increases. This behavior is opposite to that of $\nu = -1$ case as the
transition here is reversed, i.e., from layer polarized (topologically
trivial) to layer hybridized (topologically nontrivial) states.

Next, we reveal the magnetic properties of this bias driven Chern insulator.
The nonvanishing Hall conductivity implies that the symmetry of the system
supports time-reversal odd pseudovectors in the out-of-plane direction, thus a
nonzero net magnetization. However, the insulating gap protects the fully
compensated spin magnetic order~\cite{jiang_antiferromagnetic_2018}
even when the bias is applied, as is verified by Fig.~\ref{fig:altermagnet}(d) as well.
As such, the net magnetization can only be of orbital origin.
This scenario is corroborated by the calculation results shown in
Fig.~\ref{fig:orbinsulator}(a). The induced orbital magnetization depends
linearly on the bias when the latter is small, and reaches the order of $0.01$
$\mu_{\text{B}}/$nm$^{2}$ in the topologically nontrivial phase. Such a
sizable value is well within the capacity of magneto-optical experiment and is
comparable to that observed in twisted bilayer graphene~\cite{tschirhart_imaging_2021}.
Moreover, in the topologically nontrivial phase the orbital magnetization changes linearly with
respect to the chemical potential within gap (Fig.~\ref{fig:orbinsulator}(c)),
and the slope is quantified by Chern number in accordance with the St{\v{r}%
}eda formula~\cite{streda_theory_1982},
 i.e., $dM/d\mu = Ce/2\pi\hbar$, where $M$ is orbital magnetization and $C$ is Chern number.
This character offers a feasible approach to experimental
identification of Chern insulator without the need of applying magnetic field.
In Fig.~\ref{fig:orbinsulator}(a), the values of $M$ are taken at the middle of band gap at each $D$,
 and the non-differentiable points at finite $D$ are due to the change from direct to indirect band gap (c.f. the sharp turning in Fig.~\ref{fig:orbinsulator}(b)).

\emph{\color{blue}Discussion and outlook.}--In MoTe$_{2}$, the spin direction
from each valley is out of plane. For tMoTe$_2$ at filling factor $\nu=-2$, at
$D=0$, the intravalley Fock exchange induces valley coherence~\cite{liu_theories_2021},
 resulting in an in-plane spin magnetic configuration (AFMxy) as the ground state~\cite{jiang_antiferromagnetic_2018, qiu_interaction-driven_2023, liu_gate-tunable_2023}.
 The interlayer bias lifts the spin degeneracy and weakens the valley coherence,
stabilizing the AFMz state which becomes the ground state at a range of finite
$D$~\cite{jiang_antiferromagnetic_2018, liu_gate-tunable_2023}.
Starting from this AFMz phase at a finite bias (e.g., $D>0$), and upon
fast switching the bias direction, the system will go through the
neighborhood of $D=0$ where AFMxy is ground state and the AFMz state is
metastable (c.f. Fig.~\ref{fig:orbinsulator}(a) and \ref{fig:orbinsulator}(d)) Nevertheless, as the magnetic relaxation could
 be slow as compared to the electrical sweep, the system can remain in AFMz
configuration and undergo the continuous topological transition, ending at the
AFMz ground state of opposite Chern number at $D<0$, thus allowing an
electrically controlled sign switch of Chern number.

The magnetic state studied here has some interesting connections to the concept of altermagnetism
\cite{hayami_momentum-dependent_2019,shao2021spin,smejkal_beyond_2022, smejkal_emerging_2022, mazin_editorial_2022, smejkal_anomalous_2022, feng_anomalous_2022, gonzalez_betancourt_spontaneous_2023}.
For this concept,  three main features have been noted in
literature:
(1) compensated collinear spin order, (2) spin splitting band
structure or alternating spin polarization in \textit{k} space, and (3)
certain rotation symmetry connecting sublattices with opposite spin. The first
two features underly the novel response properties of altermagnets, such as
the spin-splitting torque~\cite{bai_observation_2022, bose_tilted_2022},
while (3) serves as a symmetry constraint
guaranteeing (1)~\cite{smejkal_beyond_2022, smejkal_emerging_2022, mazin_editorial_2022}.
The topological magnetic state discovered at $\nu=-2$ tMoTe$_2$
here have the first two features, but the spin-up and spin-down channels are not related by rotation symmetry.
{
Therefore, the $\nu=-2$ tMoTe$_2$ here has some similarities to but is different from the altermagnetic system.
}
The sizable orbital
magnetization here allows for efficient magnetic-field control and switch of
the two time-reversed degenerate topological states, e.g., for initialization of
a designated magnetic state with definite Chern number. With such
controllability through bias and magnetic field, the twisted TMDs bilayers
provide a promising platform for manipulations of topological magnetism.

This work is supported by Research Grant Council of Hong Kong SAR China
through grants HKU SRFS2122-7S05 and AoE/P-701/20, and the National Key R$\And$D
Program of China (2020YFA0309600).
C.X. acknowledges support by the Start-up Research Grant of University of Macau.
W.Y. also acknowledges support by the New Cornerstone Science Foundation through the XPLORER PRIZE.

\bibliographystyle{apsrev4-2}
\bibliography{orbital_Chern_insulator_tMoTe2}

\begin{thebibliography}{38}%
\makeatletter
\providecommand \@ifxundefined [1]{%
 \@ifx{#1\undefined}
}%
\providecommand \@ifnum [1]{%
 \ifnum #1\expandafter \@firstoftwo
 \else \expandafter \@secondoftwo
 \fi
}%
\providecommand \@ifx [1]{%
 \ifx #1\expandafter \@firstoftwo
 \else \expandafter \@secondoftwo
 \fi
}%
\providecommand \natexlab [1]{#1}%
\providecommand \enquote  [1]{``#1''}%
\providecommand \bibnamefont  [1]{#1}%
\providecommand \bibfnamefont [1]{#1}%
\providecommand \citenamefont [1]{#1}%
\providecommand \href@noop [0]{\@secondoftwo}%
\providecommand \href [0]{\begingroup \@sanitize@url \@href}%
\providecommand \@href[1]{\@@startlink{#1}\@@href}%
\providecommand \@@href[1]{\endgroup#1\@@endlink}%
\providecommand \@sanitize@url [0]{\catcode `\\12\catcode `\$12\catcode
  `\&12\catcode `\#12\catcode `\^12\catcode `\_12\catcode `\%12\relax}%
\providecommand \@@startlink[1]{}%
\providecommand \@@endlink[0]{}%
\providecommand \url  [0]{\begingroup\@sanitize@url \@url }%
\providecommand \@url [1]{\endgroup\@href {#1}{\urlprefix }}%
\providecommand \urlprefix  [0]{URL }%
\providecommand \Eprint [0]{\href }%
\providecommand \doibase [0]{https://doi.org/}%
\providecommand \selectlanguage [0]{\@gobble}%
\providecommand \bibinfo  [0]{\@secondoftwo}%
\providecommand \bibfield  [0]{\@secondoftwo}%
\providecommand \translation [1]{[#1]}%
\providecommand \BibitemOpen [0]{}%
\providecommand \bibitemStop [0]{}%
\providecommand \bibitemNoStop [0]{.\EOS\space}%
\providecommand \EOS [0]{\spacefactor3000\relax}%
\providecommand \BibitemShut  [1]{\csname bibitem#1\endcsname}%
\let\auto@bib@innerbib\@empty
\bibitem [{\citenamefont {Lezama}\ \emph {et~al.}(2015)\citenamefont {Lezama},
  \citenamefont {Arora}, \citenamefont {Ubaldini}, \citenamefont {Barreteau},
  \citenamefont {Giannini}, \citenamefont {Potemski},\ and\ \citenamefont
  {Morpurgo}}]{lezama_indirect--direct_2015}%
  \BibitemOpen
  \bibfield  {author} {\bibinfo {author} {\bibfnamefont {I.~G.}\ \bibnamefont
  {Lezama}}, \bibinfo {author} {\bibfnamefont {A.}~\bibnamefont {Arora}},
  \bibinfo {author} {\bibfnamefont {A.}~\bibnamefont {Ubaldini}}, \bibinfo
  {author} {\bibfnamefont {C.}~\bibnamefont {Barreteau}}, \bibinfo {author}
  {\bibfnamefont {E.}~\bibnamefont {Giannini}}, \bibinfo {author}
  {\bibfnamefont {M.}~\bibnamefont {Potemski}},\ and\ \bibinfo {author}
  {\bibfnamefont {A.~F.}\ \bibnamefont {Morpurgo}},\ }\href
  {https://doi.org/10.1021/nl5045007} {\bibfield  {journal} {\bibinfo
  {journal} {Nano Lett.}\ }\textbf {\bibinfo {volume} {15}},\ \bibinfo {pages}
  {2336} (\bibinfo {year} {2015})}\BibitemShut {NoStop}%
\bibitem [{\citenamefont {Wu}\ \emph {et~al.}(2019)\citenamefont {Wu},
  \citenamefont {Lovorn}, \citenamefont {Tutuc}, \citenamefont {Martin},\ and\
  \citenamefont {MacDonald}}]{wu_topological_2019}%
  \BibitemOpen
  \bibfield  {author} {\bibinfo {author} {\bibfnamefont {F.}~\bibnamefont
  {Wu}}, \bibinfo {author} {\bibfnamefont {T.}~\bibnamefont {Lovorn}}, \bibinfo
  {author} {\bibfnamefont {E.}~\bibnamefont {Tutuc}}, \bibinfo {author}
  {\bibfnamefont {I.}~\bibnamefont {Martin}},\ and\ \bibinfo {author}
  {\bibfnamefont {A.~H.}\ \bibnamefont {MacDonald}},\ }\href
  {https://link.aps.org/doi/10.1103/PhysRevLett.122.086402} {\bibfield
  {journal} {\bibinfo  {journal} {Phys. Rev. Lett.}\ }\textbf {\bibinfo
  {volume} {122}},\ \bibinfo {pages} {086402} (\bibinfo {year}
  {2019})}\BibitemShut {NoStop}%
\bibitem [{\citenamefont {Yu}\ \emph {et~al.}(2020)\citenamefont {Yu},
  \citenamefont {Chen},\ and\ \citenamefont {Yao}}]{yu_giant_2020}%
  \BibitemOpen
  \bibfield  {author} {\bibinfo {author} {\bibfnamefont {H.}~\bibnamefont
  {Yu}}, \bibinfo {author} {\bibfnamefont {M.}~\bibnamefont {Chen}},\ and\
  \bibinfo {author} {\bibfnamefont {W.}~\bibnamefont {Yao}},\ }\href
  {https://doi.org/10.1093/nsr/nwz117} {\bibfield  {journal} {\bibinfo
  {journal} {Natl. Sci. Rev.}\ }\textbf {\bibinfo {volume} {7}},\ \bibinfo
  {pages} {12} (\bibinfo {year} {2020})},\ \bibinfo {note} {published 13 Aug
  (2019)}\BibitemShut {NoStop}%
\bibitem [{\citenamefont {Zhai}\ and\ \citenamefont
  {Yao}(2020)}]{zhai_theory_2020}%
  \BibitemOpen
  \bibfield  {author} {\bibinfo {author} {\bibfnamefont {D.}~\bibnamefont
  {Zhai}}\ and\ \bibinfo {author} {\bibfnamefont {W.}~\bibnamefont {Yao}},\
  }\href {https://doi.org/10.1103/PhysRevMaterials.4.094002} {\bibfield
  {journal} {\bibinfo  {journal} {Phys. Rev. Mater.}\ }\textbf {\bibinfo
  {volume} {4}},\ \bibinfo {pages} {094002} (\bibinfo {year}
  {2020})}\BibitemShut {NoStop}%
\bibitem [{\citenamefont {Devakul}\ \emph {et~al.}(2021)\citenamefont
  {Devakul}, \citenamefont {Crépel}, \citenamefont {Zhang},\ and\
  \citenamefont {Fu}}]{devakul_magic_2021}%
  \BibitemOpen
  \bibfield  {author} {\bibinfo {author} {\bibfnamefont {T.}~\bibnamefont
  {Devakul}}, \bibinfo {author} {\bibfnamefont {V.}~\bibnamefont {Crépel}},
  \bibinfo {author} {\bibfnamefont {Y.}~\bibnamefont {Zhang}},\ and\ \bibinfo
  {author} {\bibfnamefont {L.}~\bibnamefont {Fu}},\ }\href
  {https://doi.org/10.1038/s41467-021-27042-9} {\bibfield  {journal} {\bibinfo
  {journal} {Nat. Commun.}\ }\textbf {\bibinfo {volume} {12}},\ \bibinfo
  {pages} {6730} (\bibinfo {year} {2021})}\BibitemShut {NoStop}%
\bibitem [{\citenamefont {Anderson}\ \emph {et~al.}(2023)\citenamefont
  {Anderson}, \citenamefont {Fan}, \citenamefont {Cai}, \citenamefont
  {Holtzmann}, \citenamefont {Taniguchi}, \citenamefont {Watanabe},
  \citenamefont {Xiao}, \citenamefont {Yao},\ and\ \citenamefont
  {Xu}}]{anderson_programming_2023}%
  \BibitemOpen
  \bibfield  {author} {\bibinfo {author} {\bibfnamefont {E.}~\bibnamefont
  {Anderson}}, \bibinfo {author} {\bibfnamefont {F.-R.}\ \bibnamefont {Fan}},
  \bibinfo {author} {\bibfnamefont {J.}~\bibnamefont {Cai}}, \bibinfo {author}
  {\bibfnamefont {W.}~\bibnamefont {Holtzmann}}, \bibinfo {author}
  {\bibfnamefont {T.}~\bibnamefont {Taniguchi}}, \bibinfo {author}
  {\bibfnamefont {K.}~\bibnamefont {Watanabe}}, \bibinfo {author}
  {\bibfnamefont {D.}~\bibnamefont {Xiao}}, \bibinfo {author} {\bibfnamefont
  {W.}~\bibnamefont {Yao}},\ and\ \bibinfo {author} {\bibfnamefont
  {X.}~\bibnamefont {Xu}},\ }\href {https://doi.org/10.1126/science.adg4268}
  {\bibfield  {journal} {\bibinfo  {journal} {Science}\ }\textbf {\bibinfo
  {volume} {381}},\ \bibinfo {pages} {325} (\bibinfo {year}
  {2023})}\BibitemShut {NoStop}%
\bibitem [{\citenamefont {Cai}\ \emph {et~al.}(2023)\citenamefont {Cai},
  \citenamefont {Anderson}, \citenamefont {Wang}, \citenamefont {Zhang},
  \citenamefont {Liu}, \citenamefont {Holtzmann}, \citenamefont {Zhang},
  \citenamefont {Fan}, \citenamefont {Taniguchi}, \citenamefont {Watanabe},
  \citenamefont {Ran}, \citenamefont {Cao}, \citenamefont {Fu}, \citenamefont
  {Xiao}, \citenamefont {Yao},\ and\ \citenamefont {Xu}}]{cai_signatures_2023}%
  \BibitemOpen
  \bibfield  {author} {\bibinfo {author} {\bibfnamefont {J.}~\bibnamefont
  {Cai}}, \bibinfo {author} {\bibfnamefont {E.}~\bibnamefont {Anderson}},
  \bibinfo {author} {\bibfnamefont {C.}~\bibnamefont {Wang}}, \bibinfo {author}
  {\bibfnamefont {X.}~\bibnamefont {Zhang}}, \bibinfo {author} {\bibfnamefont
  {X.}~\bibnamefont {Liu}}, \bibinfo {author} {\bibfnamefont {W.}~\bibnamefont
  {Holtzmann}}, \bibinfo {author} {\bibfnamefont {Y.}~\bibnamefont {Zhang}},
  \bibinfo {author} {\bibfnamefont {F.-R.}\ \bibnamefont {Fan}}, \bibinfo
  {author} {\bibfnamefont {T.}~\bibnamefont {Taniguchi}}, \bibinfo {author}
  {\bibfnamefont {K.}~\bibnamefont {Watanabe}}, \bibinfo {author}
  {\bibfnamefont {Y.}~\bibnamefont {Ran}}, \bibinfo {author} {\bibfnamefont
  {T.}~\bibnamefont {Cao}}, \bibinfo {author} {\bibfnamefont {L.}~\bibnamefont
  {Fu}}, \bibinfo {author} {\bibfnamefont {D.}~\bibnamefont {Xiao}}, \bibinfo
  {author} {\bibfnamefont {W.}~\bibnamefont {Yao}},\ and\ \bibinfo {author}
  {\bibfnamefont {X.}~\bibnamefont {Xu}},\ }\bibfield  {journal} {\bibinfo
  {journal} {Nature}\ }\href {https://doi.org/10.1038/s41586-023-06289-w}
  {10.1038/s41586-023-06289-w} (\bibinfo {year} {2023})\BibitemShut {NoStop}%
\bibitem [{\citenamefont {Zeng}\ \emph {et~al.}(2023)\citenamefont {Zeng},
  \citenamefont {Xia}, \citenamefont {Kang}, \citenamefont {Zhu}, \citenamefont
  {Knüppel}, \citenamefont {Vaswani}, \citenamefont {Watanabe}, \citenamefont
  {Taniguchi}, \citenamefont {Mak},\ and\ \citenamefont
  {Shan}}]{zeng_thermodynamic_2023}%
  \BibitemOpen
  \bibfield  {author} {\bibinfo {author} {\bibfnamefont {Y.}~\bibnamefont
  {Zeng}}, \bibinfo {author} {\bibfnamefont {Z.}~\bibnamefont {Xia}}, \bibinfo
  {author} {\bibfnamefont {K.}~\bibnamefont {Kang}}, \bibinfo {author}
  {\bibfnamefont {J.}~\bibnamefont {Zhu}}, \bibinfo {author} {\bibfnamefont
  {P.}~\bibnamefont {Knüppel}}, \bibinfo {author} {\bibfnamefont
  {C.}~\bibnamefont {Vaswani}}, \bibinfo {author} {\bibfnamefont
  {K.}~\bibnamefont {Watanabe}}, \bibinfo {author} {\bibfnamefont
  {T.}~\bibnamefont {Taniguchi}}, \bibinfo {author} {\bibfnamefont {K.~F.}\
  \bibnamefont {Mak}},\ and\ \bibinfo {author} {\bibfnamefont {J.}~\bibnamefont
  {Shan}},\ }\bibfield  {journal} {\bibinfo  {journal} {Nature}\ }\href
  {https://doi.org/10.1038/s41586-023-06452-3} {10.1038/s41586-023-06452-3}
  (\bibinfo {year} {2023})\BibitemShut {NoStop}%
\bibitem [{\citenamefont {Park}\ \emph {et~al.}(2023)\citenamefont {Park},
  \citenamefont {Cai}, \citenamefont {Anderson}, \citenamefont {Zhang},
  \citenamefont {Zhu}, \citenamefont {Liu}, \citenamefont {Wang}, \citenamefont
  {Holtzmann}, \citenamefont {Hu}, \citenamefont {Liu}, \citenamefont
  {Taniguchi}, \citenamefont {Watanabe}, \citenamefont {Chu}, \citenamefont
  {Cao}, \citenamefont {Fu}, \citenamefont {Yao}, \citenamefont {Chang},
  \citenamefont {Cobden}, \citenamefont {Xiao},\ and\ \citenamefont
  {Xu}}]{park_observation_2023}%
  \BibitemOpen
  \bibfield  {author} {\bibinfo {author} {\bibfnamefont {H.}~\bibnamefont
  {Park}}, \bibinfo {author} {\bibfnamefont {J.}~\bibnamefont {Cai}}, \bibinfo
  {author} {\bibfnamefont {E.}~\bibnamefont {Anderson}}, \bibinfo {author}
  {\bibfnamefont {Y.}~\bibnamefont {Zhang}}, \bibinfo {author} {\bibfnamefont
  {J.}~\bibnamefont {Zhu}}, \bibinfo {author} {\bibfnamefont {X.}~\bibnamefont
  {Liu}}, \bibinfo {author} {\bibfnamefont {C.}~\bibnamefont {Wang}}, \bibinfo
  {author} {\bibfnamefont {W.}~\bibnamefont {Holtzmann}}, \bibinfo {author}
  {\bibfnamefont {C.}~\bibnamefont {Hu}}, \bibinfo {author} {\bibfnamefont
  {Z.}~\bibnamefont {Liu}}, \bibinfo {author} {\bibfnamefont {T.}~\bibnamefont
  {Taniguchi}}, \bibinfo {author} {\bibfnamefont {K.}~\bibnamefont {Watanabe}},
  \bibinfo {author} {\bibfnamefont {J.-h.}\ \bibnamefont {Chu}}, \bibinfo
  {author} {\bibfnamefont {T.}~\bibnamefont {Cao}}, \bibinfo {author}
  {\bibfnamefont {L.}~\bibnamefont {Fu}}, \bibinfo {author} {\bibfnamefont
  {W.}~\bibnamefont {Yao}}, \bibinfo {author} {\bibfnamefont {C.-Z.}\
  \bibnamefont {Chang}}, \bibinfo {author} {\bibfnamefont {D.}~\bibnamefont
  {Cobden}}, \bibinfo {author} {\bibfnamefont {D.}~\bibnamefont {Xiao}},\ and\
  \bibinfo {author} {\bibfnamefont {X.}~\bibnamefont {Xu}},\ }\bibfield
  {journal} {\bibinfo  {journal} {Nature}\ }\href
  {https://doi.org/10.1038/s41586-023-06536-0} {10.1038/s41586-023-06536-0}
  (\bibinfo {year} {2023})\BibitemShut {NoStop}%
\bibitem [{\citenamefont {Foutty}\ \emph {et~al.}(2023)\citenamefont {Foutty},
  \citenamefont {Kometter}, \citenamefont {Devakul}, \citenamefont {Reddy},
  \citenamefont {Watanabe}, \citenamefont {Taniguchi}, \citenamefont {Fu},\
  and\ \citenamefont {Feldman}}]{foutty_mapping_2023}%
  \BibitemOpen
  \bibfield  {author} {\bibinfo {author} {\bibfnamefont {B.~A.}\ \bibnamefont
  {Foutty}}, \bibinfo {author} {\bibfnamefont {C.~R.}\ \bibnamefont
  {Kometter}}, \bibinfo {author} {\bibfnamefont {T.}~\bibnamefont {Devakul}},
  \bibinfo {author} {\bibfnamefont {A.~P.}\ \bibnamefont {Reddy}}, \bibinfo
  {author} {\bibfnamefont {K.}~\bibnamefont {Watanabe}}, \bibinfo {author}
  {\bibfnamefont {T.}~\bibnamefont {Taniguchi}}, \bibinfo {author}
  {\bibfnamefont {L.}~\bibnamefont {Fu}},\ and\ \bibinfo {author}
  {\bibfnamefont {B.~E.}\ \bibnamefont {Feldman}},\ }\href
  {http://arxiv.org/abs/2304.09808} {\bibfield  {journal} {\bibinfo  {journal}
  {arXiv:2304.09808}\ } (\bibinfo {year} {2023})}\BibitemShut {NoStop}%
\bibitem [{\citenamefont {Li}\ \emph {et~al.}(2021)\citenamefont {Li},
  \citenamefont {Kumar}, \citenamefont {Sun},\ and\ \citenamefont
  {Lin}}]{li_spontaneous_2021}%
  \BibitemOpen
  \bibfield  {author} {\bibinfo {author} {\bibfnamefont {H.}~\bibnamefont
  {Li}}, \bibinfo {author} {\bibfnamefont {U.}~\bibnamefont {Kumar}}, \bibinfo
  {author} {\bibfnamefont {K.}~\bibnamefont {Sun}},\ and\ \bibinfo {author}
  {\bibfnamefont {S.-Z.}\ \bibnamefont {Lin}},\ }\href
  {https://doi.org/10.1103/PhysRevResearch.3.L032070} {\bibfield  {journal}
  {\bibinfo  {journal} {Phys. Rev. Res.}\ }\textbf {\bibinfo {volume} {3}},\
  \bibinfo {pages} {L032070} (\bibinfo {year} {2021})}\BibitemShut {NoStop}%
\bibitem [{\citenamefont {Xu}\ \emph {et~al.}(2023)\citenamefont {Xu},
  \citenamefont {Sun}, \citenamefont {Jia}, \citenamefont {Liu}, \citenamefont
  {Xu}, \citenamefont {Li}, \citenamefont {Gu}, \citenamefont {Watanabe},
  \citenamefont {Taniguchi}, \citenamefont {Tong}, \citenamefont {Jia},
  \citenamefont {Shi}, \citenamefont {Jiang}, \citenamefont {Zhang},
  \citenamefont {Liu},\ and\ \citenamefont {Li}}]{xu_observation_2023}%
  \BibitemOpen
  \bibfield  {author} {\bibinfo {author} {\bibfnamefont {F.}~\bibnamefont
  {Xu}}, \bibinfo {author} {\bibfnamefont {Z.}~\bibnamefont {Sun}}, \bibinfo
  {author} {\bibfnamefont {T.}~\bibnamefont {Jia}}, \bibinfo {author}
  {\bibfnamefont {C.}~\bibnamefont {Liu}}, \bibinfo {author} {\bibfnamefont
  {C.}~\bibnamefont {Xu}}, \bibinfo {author} {\bibfnamefont {C.}~\bibnamefont
  {Li}}, \bibinfo {author} {\bibfnamefont {Y.}~\bibnamefont {Gu}}, \bibinfo
  {author} {\bibfnamefont {K.}~\bibnamefont {Watanabe}}, \bibinfo {author}
  {\bibfnamefont {T.}~\bibnamefont {Taniguchi}}, \bibinfo {author}
  {\bibfnamefont {B.}~\bibnamefont {Tong}}, \bibinfo {author} {\bibfnamefont
  {J.}~\bibnamefont {Jia}}, \bibinfo {author} {\bibfnamefont {Z.}~\bibnamefont
  {Shi}}, \bibinfo {author} {\bibfnamefont {S.}~\bibnamefont {Jiang}}, \bibinfo
  {author} {\bibfnamefont {Y.}~\bibnamefont {Zhang}}, \bibinfo {author}
  {\bibfnamefont {X.}~\bibnamefont {Liu}},\ and\ \bibinfo {author}
  {\bibfnamefont {T.}~\bibnamefont {Li}},\ }\href
  {http://arxiv.org/abs/2308.06177} {\bibfield  {journal} {\bibinfo  {journal}
  {arXiv:2308.06177}\ } (\bibinfo {year} {2023})}\BibitemShut {NoStop}%
\bibitem [{\citenamefont {Jiang}\ \emph {et~al.}(2018)\citenamefont {Jiang},
  \citenamefont {Zhou}, \citenamefont {Dai},\ and\ \citenamefont
  {Wang}}]{jiang_antiferromagnetic_2018}%
  \BibitemOpen
  \bibfield  {author} {\bibinfo {author} {\bibfnamefont {K.}~\bibnamefont
  {Jiang}}, \bibinfo {author} {\bibfnamefont {S.}~\bibnamefont {Zhou}},
  \bibinfo {author} {\bibfnamefont {X.}~\bibnamefont {Dai}},\ and\ \bibinfo
  {author} {\bibfnamefont {Z.}~\bibnamefont {Wang}},\ }\href
  {https://doi.org/10.1103/PhysRevLett.120.157205} {\bibfield  {journal}
  {\bibinfo  {journal} {Phys. Rev. Lett.}\ }\textbf {\bibinfo {volume} {120}},\
  \bibinfo {pages} {157205} (\bibinfo {year} {2018})}\BibitemShut {NoStop}%
\bibitem [{\citenamefont {Dai}\ \emph {et~al.}(2022)\citenamefont {Dai},
  \citenamefont {Li}, \citenamefont {Xu}, \citenamefont {Chen},\ and\
  \citenamefont {Xie}}]{dai_quantum_2022}%
  \BibitemOpen
  \bibfield  {author} {\bibinfo {author} {\bibfnamefont {W.-B.}\ \bibnamefont
  {Dai}}, \bibinfo {author} {\bibfnamefont {H.}~\bibnamefont {Li}}, \bibinfo
  {author} {\bibfnamefont {D.-H.}\ \bibnamefont {Xu}}, \bibinfo {author}
  {\bibfnamefont {C.-Z.}\ \bibnamefont {Chen}},\ and\ \bibinfo {author}
  {\bibfnamefont {X.~C.}\ \bibnamefont {Xie}},\ }\href
  {https://doi.org/10.1103/PhysRevB.106.245425} {\bibfield  {journal} {\bibinfo
   {journal} {Phys. Rev. B}\ }\textbf {\bibinfo {volume} {106}},\ \bibinfo
  {pages} {245425} (\bibinfo {year} {2022})}\BibitemShut {NoStop}%
\bibitem [{\citenamefont {Guo}\ \emph {et~al.}(2023)\citenamefont {Guo},
  \citenamefont {Liu},\ and\ \citenamefont {Lu}}]{guo_quantum_2023}%
  \BibitemOpen
  \bibfield  {author} {\bibinfo {author} {\bibfnamefont {P.-J.}\ \bibnamefont
  {Guo}}, \bibinfo {author} {\bibfnamefont {Z.-X.}\ \bibnamefont {Liu}},\ and\
  \bibinfo {author} {\bibfnamefont {Z.-Y.}\ \bibnamefont {Lu}},\ }\href
  {https://doi.org/10.1038/s41524-023-01025-4} {\bibfield  {journal} {\bibinfo
  {journal} {npj Comput. Mater.}\ }\textbf {\bibinfo {volume} {9}},\ \bibinfo
  {pages} {1} (\bibinfo {year} {2023})}\BibitemShut {NoStop}%
\bibitem [{\citenamefont {Hayami}\ \emph {et~al.}(2019)\citenamefont {Hayami},
  \citenamefont {Yanagi},\ and\ \citenamefont
  {Kusunose}}]{hayami_momentum-dependent_2019}%
  \BibitemOpen
  \bibfield  {author} {\bibinfo {author} {\bibfnamefont {S.}~\bibnamefont
  {Hayami}}, \bibinfo {author} {\bibfnamefont {Y.}~\bibnamefont {Yanagi}},\
  and\ \bibinfo {author} {\bibfnamefont {H.}~\bibnamefont {Kusunose}},\ }\href
  {https://doi.org/10.7566/JPSJ.88.123702} {\bibfield  {journal} {\bibinfo
  {journal} {J. Phys. Soc. Jpn.}\ }\textbf {\bibinfo {volume} {88}},\ \bibinfo
  {pages} {123702} (\bibinfo {year} {2019})}\BibitemShut {NoStop}%
\bibitem [{\citenamefont {Shao}\ \emph {et~al.}(2021)\citenamefont {Shao},
  \citenamefont {Zhang}, \citenamefont {Li}, \citenamefont {Eom},\ and\
  \citenamefont {Tsymbal}}]{shao2021spin}%
  \BibitemOpen
  \bibfield  {author} {\bibinfo {author} {\bibfnamefont {D.-F.}\ \bibnamefont
  {Shao}}, \bibinfo {author} {\bibfnamefont {S.-H.}\ \bibnamefont {Zhang}},
  \bibinfo {author} {\bibfnamefont {M.}~\bibnamefont {Li}}, \bibinfo {author}
  {\bibfnamefont {C.-B.}\ \bibnamefont {Eom}},\ and\ \bibinfo {author}
  {\bibfnamefont {E.~Y.}\ \bibnamefont {Tsymbal}},\ }\href
  {https://doi.org/10.1038/s41467-021-26915-3} {\bibfield  {journal} {\bibinfo
  {journal} {Nat. Commun.}\ }\textbf {\bibinfo {volume} {12}},\ \bibinfo
  {pages} {7061} (\bibinfo {year} {2021})}\BibitemShut {NoStop}%
\bibitem [{\citenamefont {Smejkal}\ \emph
  {et~al.}(2022{\natexlab{a}})\citenamefont {Smejkal}, \citenamefont {Sinova},\
  and\ \citenamefont {Jungwirth}}]{smejkal_beyond_2022}%
  \BibitemOpen
  \bibfield  {author} {\bibinfo {author} {\bibfnamefont {L.}~\bibnamefont
  {Smejkal}}, \bibinfo {author} {\bibfnamefont {J.}~\bibnamefont {Sinova}},\
  and\ \bibinfo {author} {\bibfnamefont {T.}~\bibnamefont {Jungwirth}},\ }\href
  {https://doi.org/10.1103/PhysRevX.12.031042} {\bibfield  {journal} {\bibinfo
  {journal} {Phys. Rev. X}\ }\textbf {\bibinfo {volume} {12}},\ \bibinfo
  {pages} {031042} (\bibinfo {year} {2022}{\natexlab{a}})}\BibitemShut
  {NoStop}%
\bibitem [{\citenamefont {Smejkal}\ \emph
  {et~al.}(2022{\natexlab{b}})\citenamefont {Smejkal}, \citenamefont {Sinova},\
  and\ \citenamefont {Jungwirth}}]{smejkal_emerging_2022}%
  \BibitemOpen
  \bibfield  {author} {\bibinfo {author} {\bibfnamefont {L.}~\bibnamefont
  {Smejkal}}, \bibinfo {author} {\bibfnamefont {J.}~\bibnamefont {Sinova}},\
  and\ \bibinfo {author} {\bibfnamefont {T.}~\bibnamefont {Jungwirth}},\ }\href
  {https://doi.org/10.1103/PhysRevX.12.040501} {\bibfield  {journal} {\bibinfo
  {journal} {Phys. Rev. X}\ }\textbf {\bibinfo {volume} {12}},\ \bibinfo
  {pages} {040501} (\bibinfo {year} {2022}{\natexlab{b}})}\BibitemShut
  {NoStop}%
\bibitem [{\citenamefont {Mazin}\ and\ \citenamefont {{The PRX
  Editors}}(2022)}]{mazin_editorial_2022}%
  \BibitemOpen
  \bibfield  {author} {\bibinfo {author} {\bibfnamefont {I.}~\bibnamefont
  {Mazin}}\ and\ \bibinfo {author} {\bibnamefont {{The PRX Editors}}},\ }\href
  {https://doi.org/10.1103/PhysRevX.12.040002} {\bibfield  {journal} {\bibinfo
  {journal} {Phys. Rev. X}\ }\textbf {\bibinfo {volume} {12}},\ \bibinfo
  {pages} {040002} (\bibinfo {year} {2022})}\BibitemShut {NoStop}%
\bibitem [{\citenamefont {Šmejkal}\ \emph {et~al.}(2022)\citenamefont
  {Šmejkal}, \citenamefont {MacDonald}, \citenamefont {Sinova}, \citenamefont
  {Nakatsuji},\ and\ \citenamefont {Jungwirth}}]{smejkal_anomalous_2022}%
  \BibitemOpen
  \bibfield  {author} {\bibinfo {author} {\bibfnamefont {L.}~\bibnamefont
  {Šmejkal}}, \bibinfo {author} {\bibfnamefont {A.~H.}\ \bibnamefont
  {MacDonald}}, \bibinfo {author} {\bibfnamefont {J.}~\bibnamefont {Sinova}},
  \bibinfo {author} {\bibfnamefont {S.}~\bibnamefont {Nakatsuji}},\ and\
  \bibinfo {author} {\bibfnamefont {T.}~\bibnamefont {Jungwirth}},\ }\href
  {https://doi.org/10.1038/s41578-022-00430-3} {\bibfield  {journal} {\bibinfo
  {journal} {Nat. Rev. Mater.}\ }\textbf {\bibinfo {volume} {7}},\ \bibinfo
  {pages} {482} (\bibinfo {year} {2022})}\BibitemShut {NoStop}%
\bibitem [{\citenamefont {Feng}\ \emph {et~al.}(2022)\citenamefont {Feng},
  \citenamefont {Zhou}, \citenamefont {Šmejkal}, \citenamefont {Wu},
  \citenamefont {Zhu}, \citenamefont {Guo}, \citenamefont
  {González-Hernández}, \citenamefont {Wang}, \citenamefont {Yan},
  \citenamefont {Qin}, \citenamefont {Zhang}, \citenamefont {Wu}, \citenamefont
  {Chen}, \citenamefont {Meng}, \citenamefont {Liu}, \citenamefont {Xia},
  \citenamefont {Sinova}, \citenamefont {Jungwirth},\ and\ \citenamefont
  {Liu}}]{feng_anomalous_2022}%
  \BibitemOpen
  \bibfield  {author} {\bibinfo {author} {\bibfnamefont {Z.}~\bibnamefont
  {Feng}}, \bibinfo {author} {\bibfnamefont {X.}~\bibnamefont {Zhou}}, \bibinfo
  {author} {\bibfnamefont {L.}~\bibnamefont {Šmejkal}}, \bibinfo {author}
  {\bibfnamefont {L.}~\bibnamefont {Wu}}, \bibinfo {author} {\bibfnamefont
  {Z.}~\bibnamefont {Zhu}}, \bibinfo {author} {\bibfnamefont {H.}~\bibnamefont
  {Guo}}, \bibinfo {author} {\bibfnamefont {R.}~\bibnamefont
  {González-Hernández}}, \bibinfo {author} {\bibfnamefont {X.}~\bibnamefont
  {Wang}}, \bibinfo {author} {\bibfnamefont {H.}~\bibnamefont {Yan}}, \bibinfo
  {author} {\bibfnamefont {P.}~\bibnamefont {Qin}}, \bibinfo {author}
  {\bibfnamefont {X.}~\bibnamefont {Zhang}}, \bibinfo {author} {\bibfnamefont
  {H.}~\bibnamefont {Wu}}, \bibinfo {author} {\bibfnamefont {H.}~\bibnamefont
  {Chen}}, \bibinfo {author} {\bibfnamefont {Z.}~\bibnamefont {Meng}}, \bibinfo
  {author} {\bibfnamefont {L.}~\bibnamefont {Liu}}, \bibinfo {author}
  {\bibfnamefont {Z.}~\bibnamefont {Xia}}, \bibinfo {author} {\bibfnamefont
  {J.}~\bibnamefont {Sinova}}, \bibinfo {author} {\bibfnamefont
  {T.}~\bibnamefont {Jungwirth}},\ and\ \bibinfo {author} {\bibfnamefont
  {Z.}~\bibnamefont {Liu}},\ }\href
  {https://doi.org/10.1038/s41928-022-00866-z} {\bibfield  {journal} {\bibinfo
  {journal} {Nat. Electron.}\ }\textbf {\bibinfo {volume} {5}},\ \bibinfo
  {pages} {735} (\bibinfo {year} {2022})}\BibitemShut {NoStop}%
\bibitem [{\citenamefont {Gonzalez~Betancourt}\ \emph
  {et~al.}(2023)\citenamefont {Gonzalez~Betancourt}, \citenamefont {Zubac},
  \citenamefont {Gonzalez-Hernandez}, \citenamefont {Geishendorf},
  \citenamefont {Soban}, \citenamefont {Springholz}, \citenamefont {Olejnik},
  \citenamefont {Smejkal}, \citenamefont {Sinova}, \citenamefont {Jungwirth},
  \citenamefont {Goennenwein}, \citenamefont {Thomas}, \citenamefont
  {Reichlova}, \citenamefont {Zelezny},\ and\ \citenamefont
  {Kriegner}}]{gonzalez_betancourt_spontaneous_2023}%
  \BibitemOpen
  \bibfield  {author} {\bibinfo {author} {\bibfnamefont {R.~D.}\ \bibnamefont
  {Gonzalez~Betancourt}}, \bibinfo {author} {\bibfnamefont {J.}~\bibnamefont
  {Zubac}}, \bibinfo {author} {\bibfnamefont {R.}~\bibnamefont
  {Gonzalez-Hernandez}}, \bibinfo {author} {\bibfnamefont {K.}~\bibnamefont
  {Geishendorf}}, \bibinfo {author} {\bibfnamefont {Z.}~\bibnamefont {Soban}},
  \bibinfo {author} {\bibfnamefont {G.}~\bibnamefont {Springholz}}, \bibinfo
  {author} {\bibfnamefont {K.}~\bibnamefont {Olejnik}}, \bibinfo {author}
  {\bibfnamefont {L.}~\bibnamefont {Smejkal}}, \bibinfo {author} {\bibfnamefont
  {J.}~\bibnamefont {Sinova}}, \bibinfo {author} {\bibfnamefont
  {T.}~\bibnamefont {Jungwirth}}, \bibinfo {author} {\bibfnamefont {S.~T.~B.}\
  \bibnamefont {Goennenwein}}, \bibinfo {author} {\bibfnamefont
  {A.}~\bibnamefont {Thomas}}, \bibinfo {author} {\bibfnamefont
  {H.}~\bibnamefont {Reichlova}}, \bibinfo {author} {\bibfnamefont
  {J.}~\bibnamefont {Zelezny}},\ and\ \bibinfo {author} {\bibfnamefont
  {D.}~\bibnamefont {Kriegner}},\ }\href
  {https://doi.org/10.1103/PhysRevLett.130.036702} {\bibfield  {journal}
  {\bibinfo  {journal} {Phys. Rev. Lett.}\ }\textbf {\bibinfo {volume} {130}},\
  \bibinfo {pages} {036702} (\bibinfo {year} {2023})}\BibitemShut {NoStop}%
\bibitem [{\citenamefont {Streda}(1982)}]{streda_theory_1982}%
  \BibitemOpen
  \bibfield  {author} {\bibinfo {author} {\bibfnamefont {P.}~\bibnamefont
  {Streda}},\ }\href {https://doi.org/10.1088/0022-3719/15/22/005} {\bibfield
  {journal} {\bibinfo  {journal} {J. of Phys. C: Solid State Phys.}\ }\textbf
  {\bibinfo {volume} {15}},\ \bibinfo {pages} {L717} (\bibinfo {year}
  {1982})}\BibitemShut {NoStop}%
\bibitem [{\citenamefont {Zhu}\ \emph {et~al.}(2020)\citenamefont {Zhu},
  \citenamefont {Su},\ and\ \citenamefont
  {MacDonald}}]{zhu_voltage-controlled_2020}%
  \BibitemOpen
  \bibfield  {author} {\bibinfo {author} {\bibfnamefont {J.}~\bibnamefont
  {Zhu}}, \bibinfo {author} {\bibfnamefont {J.-J.}\ \bibnamefont {Su}},\ and\
  \bibinfo {author} {\bibfnamefont {A.~H.}\ \bibnamefont {MacDonald}},\ }\href
  {https://doi.org/10.1103/PhysRevLett.125.227702} {\bibfield  {journal}
  {\bibinfo  {journal} {Phys. Rev. Lett.}\ }\textbf {\bibinfo {volume} {125}},\
  \bibinfo {pages} {227702} (\bibinfo {year} {2020})}\BibitemShut {NoStop}%
\bibitem [{\citenamefont {Sharpe}\ \emph {et~al.}(2019)\citenamefont {Sharpe},
  \citenamefont {Fox}, \citenamefont {Barnard}, \citenamefont {Finney},
  \citenamefont {Watanabe}, \citenamefont {Taniguchi}, \citenamefont
  {Kastner},\ and\ \citenamefont {Goldhaber-Gordon}}]{sharpe_emergent_2019}%
  \BibitemOpen
  \bibfield  {author} {\bibinfo {author} {\bibfnamefont {A.~L.}\ \bibnamefont
  {Sharpe}}, \bibinfo {author} {\bibfnamefont {E.~J.}\ \bibnamefont {Fox}},
  \bibinfo {author} {\bibfnamefont {A.~W.}\ \bibnamefont {Barnard}}, \bibinfo
  {author} {\bibfnamefont {J.}~\bibnamefont {Finney}}, \bibinfo {author}
  {\bibfnamefont {K.}~\bibnamefont {Watanabe}}, \bibinfo {author}
  {\bibfnamefont {T.}~\bibnamefont {Taniguchi}}, \bibinfo {author}
  {\bibfnamefont {M.~A.}\ \bibnamefont {Kastner}},\ and\ \bibinfo {author}
  {\bibfnamefont {D.}~\bibnamefont {Goldhaber-Gordon}},\ }\href
  {https://doi.org/10.1126/science.aaw3780} {\bibfield  {journal} {\bibinfo
  {journal} {Science}\ }\textbf {\bibinfo {volume} {365}},\ \bibinfo {pages}
  {605} (\bibinfo {year} {2019})}\BibitemShut {NoStop}%
\bibitem [{\citenamefont {Serlin}\ \emph {et~al.}(2020)\citenamefont {Serlin},
  \citenamefont {Tschirhart}, \citenamefont {Polshyn}, \citenamefont {Zhang},
  \citenamefont {Zhu}, \citenamefont {Watanabe}, \citenamefont {Taniguchi},
  \citenamefont {Balents},\ and\ \citenamefont
  {Young}}]{serlin_intrinsic_2020}%
  \BibitemOpen
  \bibfield  {author} {\bibinfo {author} {\bibfnamefont {M.}~\bibnamefont
  {Serlin}}, \bibinfo {author} {\bibfnamefont {C.~L.}\ \bibnamefont
  {Tschirhart}}, \bibinfo {author} {\bibfnamefont {H.}~\bibnamefont {Polshyn}},
  \bibinfo {author} {\bibfnamefont {Y.}~\bibnamefont {Zhang}}, \bibinfo
  {author} {\bibfnamefont {J.}~\bibnamefont {Zhu}}, \bibinfo {author}
  {\bibfnamefont {K.}~\bibnamefont {Watanabe}}, \bibinfo {author}
  {\bibfnamefont {T.}~\bibnamefont {Taniguchi}}, \bibinfo {author}
  {\bibfnamefont {L.}~\bibnamefont {Balents}},\ and\ \bibinfo {author}
  {\bibfnamefont {A.~F.}\ \bibnamefont {Young}},\ }\href
  {https://doi.org/10.1126/science.aay5533} {\bibfield  {journal} {\bibinfo
  {journal} {Science}\ }\textbf {\bibinfo {volume} {367}},\ \bibinfo {pages}
  {900} (\bibinfo {year} {2020})}\BibitemShut {NoStop}%
\bibitem [{\citenamefont {Polshyn}\ \emph {et~al.}(2020)\citenamefont
  {Polshyn}, \citenamefont {Zhu}, \citenamefont {Kumar}, \citenamefont {Zhang},
  \citenamefont {Yang}, \citenamefont {Tschirhart}, \citenamefont {Serlin},
  \citenamefont {Watanabe}, \citenamefont {Taniguchi}, \citenamefont
  {MacDonald},\ and\ \citenamefont {Young}}]{polshyn_electrical_2020}%
  \BibitemOpen
  \bibfield  {author} {\bibinfo {author} {\bibfnamefont {H.}~\bibnamefont
  {Polshyn}}, \bibinfo {author} {\bibfnamefont {J.}~\bibnamefont {Zhu}},
  \bibinfo {author} {\bibfnamefont {M.~A.}\ \bibnamefont {Kumar}}, \bibinfo
  {author} {\bibfnamefont {Y.}~\bibnamefont {Zhang}}, \bibinfo {author}
  {\bibfnamefont {F.}~\bibnamefont {Yang}}, \bibinfo {author} {\bibfnamefont
  {C.~L.}\ \bibnamefont {Tschirhart}}, \bibinfo {author} {\bibfnamefont
  {M.}~\bibnamefont {Serlin}}, \bibinfo {author} {\bibfnamefont
  {K.}~\bibnamefont {Watanabe}}, \bibinfo {author} {\bibfnamefont
  {T.}~\bibnamefont {Taniguchi}}, \bibinfo {author} {\bibfnamefont {A.~H.}\
  \bibnamefont {MacDonald}},\ and\ \bibinfo {author} {\bibfnamefont {A.~F.}\
  \bibnamefont {Young}},\ }\href {https://doi.org/10.1038/s41586-020-2963-8}
  {\bibfield  {journal} {\bibinfo  {journal} {Nature}\ }\textbf {\bibinfo
  {volume} {588}},\ \bibinfo {pages} {66} (\bibinfo {year} {2020})}\BibitemShut
  {NoStop}%
\bibitem [{\citenamefont {Xie}\ \emph {et~al.}(2021)\citenamefont {Xie},
  \citenamefont {Pierce}, \citenamefont {Park}, \citenamefont {Parker},
  \citenamefont {Khalaf}, \citenamefont {Ledwith}, \citenamefont {Cao},
  \citenamefont {Lee}, \citenamefont {Chen}, \citenamefont {Forrester},
  \citenamefont {Watanabe}, \citenamefont {Taniguchi}, \citenamefont
  {Vishwanath}, \citenamefont {Jarillo-Herrero},\ and\ \citenamefont
  {Yacoby}}]{xie_fractional_2021}%
  \BibitemOpen
  \bibfield  {author} {\bibinfo {author} {\bibfnamefont {Y.}~\bibnamefont
  {Xie}}, \bibinfo {author} {\bibfnamefont {A.~T.}\ \bibnamefont {Pierce}},
  \bibinfo {author} {\bibfnamefont {J.~M.}\ \bibnamefont {Park}}, \bibinfo
  {author} {\bibfnamefont {D.~E.}\ \bibnamefont {Parker}}, \bibinfo {author}
  {\bibfnamefont {E.}~\bibnamefont {Khalaf}}, \bibinfo {author} {\bibfnamefont
  {P.}~\bibnamefont {Ledwith}}, \bibinfo {author} {\bibfnamefont
  {Y.}~\bibnamefont {Cao}}, \bibinfo {author} {\bibfnamefont {S.~H.}\
  \bibnamefont {Lee}}, \bibinfo {author} {\bibfnamefont {S.}~\bibnamefont
  {Chen}}, \bibinfo {author} {\bibfnamefont {P.~R.}\ \bibnamefont {Forrester}},
  \bibinfo {author} {\bibfnamefont {K.}~\bibnamefont {Watanabe}}, \bibinfo
  {author} {\bibfnamefont {T.}~\bibnamefont {Taniguchi}}, \bibinfo {author}
  {\bibfnamefont {A.}~\bibnamefont {Vishwanath}}, \bibinfo {author}
  {\bibfnamefont {P.}~\bibnamefont {Jarillo-Herrero}},\ and\ \bibinfo {author}
  {\bibfnamefont {A.}~\bibnamefont {Yacoby}},\ }\href
  {https://doi.org/10.1038/s41586-021-04002-3} {\bibfield  {journal} {\bibinfo
  {journal} {Nature}\ }\textbf {\bibinfo {volume} {600}},\ \bibinfo {pages}
  {439} (\bibinfo {year} {2021})}\BibitemShut {NoStop}%
\bibitem [{\citenamefont {Wang}\ \emph {et~al.}(2023)\citenamefont {Wang},
  \citenamefont {Zhang}, \citenamefont {Liu}, \citenamefont {He}, \citenamefont
  {Xu}, \citenamefont {Ran}, \citenamefont {Cao},\ and\ \citenamefont
  {Xiao}}]{wang_fractional_2023}%
  \BibitemOpen
  \bibfield  {author} {\bibinfo {author} {\bibfnamefont {C.}~\bibnamefont
  {Wang}}, \bibinfo {author} {\bibfnamefont {X.-W.}\ \bibnamefont {Zhang}},
  \bibinfo {author} {\bibfnamefont {X.}~\bibnamefont {Liu}}, \bibinfo {author}
  {\bibfnamefont {Y.}~\bibnamefont {He}}, \bibinfo {author} {\bibfnamefont
  {X.}~\bibnamefont {Xu}}, \bibinfo {author} {\bibfnamefont {Y.}~\bibnamefont
  {Ran}}, \bibinfo {author} {\bibfnamefont {T.}~\bibnamefont {Cao}},\ and\
  \bibinfo {author} {\bibfnamefont {D.}~\bibnamefont {Xiao}},\ }\href
  {http://arxiv.org/abs/2304.11864} {\bibfield  {journal} {\bibinfo  {journal}
  {arXiv:2304.11864}\ } (\bibinfo {year} {2023})}\BibitemShut {NoStop}%
\bibitem [{\citenamefont {Hohenadler}\ \emph {et~al.}(2012)\citenamefont
  {Hohenadler}, \citenamefont {Meng}, \citenamefont {Lang}, \citenamefont
  {Wessel}, \citenamefont {Muramatsu},\ and\ \citenamefont
  {Assaad}}]{hohenadler_quantum_2012}%
  \BibitemOpen
  \bibfield  {author} {\bibinfo {author} {\bibfnamefont {M.}~\bibnamefont
  {Hohenadler}}, \bibinfo {author} {\bibfnamefont {Z.~Y.}\ \bibnamefont
  {Meng}}, \bibinfo {author} {\bibfnamefont {T.~C.}\ \bibnamefont {Lang}},
  \bibinfo {author} {\bibfnamefont {S.}~\bibnamefont {Wessel}}, \bibinfo
  {author} {\bibfnamefont {A.}~\bibnamefont {Muramatsu}},\ and\ \bibinfo
  {author} {\bibfnamefont {F.~F.}\ \bibnamefont {Assaad}},\ }\href
  {https://doi.org/10.1103/PhysRevB.85.115132} {\bibfield  {journal} {\bibinfo
  {journal} {Phys. Rev. B}\ }\textbf {\bibinfo {volume} {85}},\ \bibinfo
  {pages} {115132} (\bibinfo {year} {2012})}\BibitemShut {NoStop}%
\bibitem [{\citenamefont {Qiu}\ \emph {et~al.}(2023)\citenamefont {Qiu},
  \citenamefont {Li}, \citenamefont {Luo},\ and\ \citenamefont
  {Wu}}]{qiu_interaction-driven_2023}%
  \BibitemOpen
  \bibfield  {author} {\bibinfo {author} {\bibfnamefont {W.-X.}\ \bibnamefont
  {Qiu}}, \bibinfo {author} {\bibfnamefont {B.}~\bibnamefont {Li}}, \bibinfo
  {author} {\bibfnamefont {X.-J.}\ \bibnamefont {Luo}},\ and\ \bibinfo {author}
  {\bibfnamefont {F.}~\bibnamefont {Wu}},\ }\href
  {http://arxiv.org/abs/2305.01006} {\bibfield  {journal} {\bibinfo  {journal}
  {arXiv:2305.01006}\ } (\bibinfo {year} {2023})}\BibitemShut {NoStop}%
\bibitem [{\citenamefont {Liu}\ \emph {et~al.}(2023)\citenamefont {Liu},
  \citenamefont {Wang}, \citenamefont {Zhang}, \citenamefont {Cao},\ and\
  \citenamefont {Xiao}}]{liu_gate-tunable_2023}%
  \BibitemOpen
  \bibfield  {author} {\bibinfo {author} {\bibfnamefont {X.}~\bibnamefont
  {Liu}}, \bibinfo {author} {\bibfnamefont {C.}~\bibnamefont {Wang}}, \bibinfo
  {author} {\bibfnamefont {X.-W.}\ \bibnamefont {Zhang}}, \bibinfo {author}
  {\bibfnamefont {T.}~\bibnamefont {Cao}},\ and\ \bibinfo {author}
  {\bibfnamefont {D.}~\bibnamefont {Xiao}},\ }\href
  {http://arxiv.org/abs/2308.07488} {\bibfield  {journal} {\bibinfo  {journal}
  {arXiv:2308.07488}\ } (\bibinfo {year} {2023})}\BibitemShut {NoStop}%
\bibitem [{\citenamefont {Xiao}\ \emph {et~al.}(2010)\citenamefont {Xiao},
  \citenamefont {Chang},\ and\ \citenamefont {Niu}}]{xiao_berry_2010}%
  \BibitemOpen
  \bibfield  {author} {\bibinfo {author} {\bibfnamefont {D.}~\bibnamefont
  {Xiao}}, \bibinfo {author} {\bibfnamefont {M.-C.}\ \bibnamefont {Chang}},\
  and\ \bibinfo {author} {\bibfnamefont {Q.}~\bibnamefont {Niu}},\ }\href
  {https://doi.org/10.1103/RevModPhys.82.1959} {\bibfield  {journal} {\bibinfo
  {journal} {Rev. of Mod. Phys.}\ }\textbf {\bibinfo {volume} {82}},\ \bibinfo
  {pages} {1959} (\bibinfo {year} {2010})}\BibitemShut {NoStop}%
\bibitem [{\citenamefont {Tschirhart}\ \emph {et~al.}(2021)\citenamefont
  {Tschirhart}, \citenamefont {Serlin}, \citenamefont {Polshyn}, \citenamefont
  {Shragai}, \citenamefont {Xia}, \citenamefont {Zhu}, \citenamefont {Zhang},
  \citenamefont {Watanabe}, \citenamefont {Taniguchi}, \citenamefont {Huber},\
  and\ \citenamefont {Young}}]{tschirhart_imaging_2021}%
  \BibitemOpen
  \bibfield  {author} {\bibinfo {author} {\bibfnamefont {C.~L.}\ \bibnamefont
  {Tschirhart}}, \bibinfo {author} {\bibfnamefont {M.}~\bibnamefont {Serlin}},
  \bibinfo {author} {\bibfnamefont {H.}~\bibnamefont {Polshyn}}, \bibinfo
  {author} {\bibfnamefont {A.}~\bibnamefont {Shragai}}, \bibinfo {author}
  {\bibfnamefont {Z.}~\bibnamefont {Xia}}, \bibinfo {author} {\bibfnamefont
  {J.}~\bibnamefont {Zhu}}, \bibinfo {author} {\bibfnamefont {Y.}~\bibnamefont
  {Zhang}}, \bibinfo {author} {\bibfnamefont {K.}~\bibnamefont {Watanabe}},
  \bibinfo {author} {\bibfnamefont {T.}~\bibnamefont {Taniguchi}}, \bibinfo
  {author} {\bibfnamefont {M.~E.}\ \bibnamefont {Huber}},\ and\ \bibinfo
  {author} {\bibfnamefont {A.~F.}\ \bibnamefont {Young}},\ }\href
  {https://doi.org/10.1126/science.abd3190} {\bibfield  {journal} {\bibinfo
  {journal} {Science}\ }\textbf {\bibinfo {volume} {372}},\ \bibinfo {pages}
  {1323} (\bibinfo {year} {2021})}\BibitemShut {NoStop}%
\bibitem [{\citenamefont {Liu}\ and\ \citenamefont
  {Dai}(2021)}]{liu_theories_2021}%
  \BibitemOpen
  \bibfield  {author} {\bibinfo {author} {\bibfnamefont {J.}~\bibnamefont
  {Liu}}\ and\ \bibinfo {author} {\bibfnamefont {X.}~\bibnamefont {Dai}},\
  }\href {https://doi.org/10.1103/PhysRevB.103.035427} {\bibfield  {journal}
  {\bibinfo  {journal} {Phys. Rev. B}\ }\textbf {\bibinfo {volume} {103}},\
  \bibinfo {pages} {035427} (\bibinfo {year} {2021})}\BibitemShut {NoStop}%
\bibitem [{\citenamefont {Bai}\ \emph {et~al.}(2022)\citenamefont {Bai},
  \citenamefont {Han}, \citenamefont {Feng}, \citenamefont {Zhou},
  \citenamefont {Su}, \citenamefont {Wang}, \citenamefont {Liao}, \citenamefont
  {Zhu}, \citenamefont {Chen}, \citenamefont {Pan}, \citenamefont {Fan},\ and\
  \citenamefont {Song}}]{bai_observation_2022}%
  \BibitemOpen
  \bibfield  {author} {\bibinfo {author} {\bibfnamefont {H.}~\bibnamefont
  {Bai}}, \bibinfo {author} {\bibfnamefont {L.}~\bibnamefont {Han}}, \bibinfo
  {author} {\bibfnamefont {X.~Y.}\ \bibnamefont {Feng}}, \bibinfo {author}
  {\bibfnamefont {Y.~J.}\ \bibnamefont {Zhou}}, \bibinfo {author}
  {\bibfnamefont {R.~X.}\ \bibnamefont {Su}}, \bibinfo {author} {\bibfnamefont
  {Q.}~\bibnamefont {Wang}}, \bibinfo {author} {\bibfnamefont {L.~Y.}\
  \bibnamefont {Liao}}, \bibinfo {author} {\bibfnamefont {W.~X.}\ \bibnamefont
  {Zhu}}, \bibinfo {author} {\bibfnamefont {X.~Z.}\ \bibnamefont {Chen}},
  \bibinfo {author} {\bibfnamefont {F.}~\bibnamefont {Pan}}, \bibinfo {author}
  {\bibfnamefont {X.~L.}\ \bibnamefont {Fan}},\ and\ \bibinfo {author}
  {\bibfnamefont {C.}~\bibnamefont {Song}},\ }\href
  {https://doi.org/10.1103/PhysRevLett.128.197202} {\bibfield  {journal}
  {\bibinfo  {journal} {Phys. Rev. Lett.}\ }\textbf {\bibinfo {volume} {128}},\
  \bibinfo {pages} {197202} (\bibinfo {year} {2022})}\BibitemShut {NoStop}%
\bibitem [{\citenamefont {Bose}\ \emph {et~al.}(2022)\citenamefont {Bose},
  \citenamefont {Schreiber}, \citenamefont {Jain}, \citenamefont {Shao},
  \citenamefont {Nair}, \citenamefont {Sun}, \citenamefont {Zhang},
  \citenamefont {Muller}, \citenamefont {Tsymbal}, \citenamefont {Schlom},\
  and\ \citenamefont {Ralph}}]{bose_tilted_2022}%
  \BibitemOpen
  \bibfield  {author} {\bibinfo {author} {\bibfnamefont {A.}~\bibnamefont
  {Bose}}, \bibinfo {author} {\bibfnamefont {N.~J.}\ \bibnamefont {Schreiber}},
  \bibinfo {author} {\bibfnamefont {R.}~\bibnamefont {Jain}}, \bibinfo {author}
  {\bibfnamefont {D.-F.}\ \bibnamefont {Shao}}, \bibinfo {author}
  {\bibfnamefont {H.~P.}\ \bibnamefont {Nair}}, \bibinfo {author}
  {\bibfnamefont {J.}~\bibnamefont {Sun}}, \bibinfo {author} {\bibfnamefont
  {X.~S.}\ \bibnamefont {Zhang}}, \bibinfo {author} {\bibfnamefont {D.~A.}\
  \bibnamefont {Muller}}, \bibinfo {author} {\bibfnamefont {E.~Y.}\
  \bibnamefont {Tsymbal}}, \bibinfo {author} {\bibfnamefont {D.~G.}\
  \bibnamefont {Schlom}},\ and\ \bibinfo {author} {\bibfnamefont {D.~C.}\
  \bibnamefont {Ralph}},\ }\href {https://doi.org/10.1038/s41928-022-00744-8}
  {\bibfield  {journal} {\bibinfo  {journal} {Nat. Electron.}\ }\textbf
  {\bibinfo {volume} {5}},\ \bibinfo {pages} {267} (\bibinfo {year}
  {2022})}\BibitemShut {NoStop}%
\end{thebibliography}%

\end{document}